\documentclass[amsmath,amssymb,aps,prb,reprint,twocolumn,superscriptaddress,longbibliography,floatfix]{revtex4-2}

\usepackage{amsmath}
\usepackage{amsfonts}
\usepackage{graphicx}
\usepackage{xcolor}
\usepackage{ulem}
\usepackage{units}
\usepackage{float}
\usepackage{xspace}
\usepackage{subfigure}
\usepackage{caption}
\usepackage{subcaption}
\usepackage{hyperref}
\usepackage{braket}
\usepackage{url}
\usepackage{booktabs}
\usepackage{multirow}
\usepackage{lipsum}
\begin{document}

\title{Competing Localizations on Disordered Non-Hermitian Random Graph Lattice}

\author{S.~Rahul}
\affiliation{Department of Science and Humanities, PES University EC Campus, Bangalore 560100, India}

\author{A.~Harshitha}
\affiliation{Department of Mathematics, Manipal Institute of Technology Bengaluru,\\ Manipal Academy of Higher Education, Manipal, India}

\begin{abstract}
Phase transitions in one-dimensional lattice systems are well established and have been extensively studied within both Hermitian and non-Hermitian frameworks. In this work, we extend this understanding to a more general setting by investigating localization and delocalization transitions and the behavior of the non-Hermitian skin effect (NHSE) using a tight-binding model on a generalized random graph lattice. Our model incorporates three key parameters—asymmetric hopping $\Delta$, on-site disorder $W$, and a random long-range coupling $p$ that together define the underlying random graph structure. By varying $p$, $\Delta$, and the disorder strength, we explore the interplay between topology, randomness, and non-Hermiticity in determining localization properties. Our results show a strong competition between skin effect driven and Anderson driven localizations across parameter regimes. Notably, even in the presence of strong disorder, skin effect driven localization coexists with Anderson driven localization. We further discuss the relevance of these results to machine-learning architectures and information propagation in complex networks and other real world problems.
\end{abstract}

\maketitle

\section*{Introduction}

Understanding the interplay between disorder, non-Hermiticity, and network topology has become a central problem in contemporary condensed matter and complex systems physics. While localization phenomena such as Anderson localization and the non-Hermitian skin effect are well understood in one-dimensional lattice models \cite{PhysRevE.109.044315,yuce2016majorana,wang2015spontaneous,bergholtz2019exceptional,gong2018topological,lieu2018topological,klett2017relation,PhysRevB.31.2437,PhysRevLett.110.176403,Zeng_2022,PhysRevB.106.014207,Wang_2021}, real physical systems—ranging from photonic and acoustic lattices with gain–loss imbalance to biological, neural, and active-matter networks—are often characterized by irregular, directed, and disordered connectivity \cite{estrada2013graph,PhysRevB.111.054116,PhysRevResearch.7.023132,kleftogiannis2024quantum}. Extending non-Hermitian tight-binding models to random graphs provides a natural framework to explore how disorder induced (Anderson) and non-reciprocity induced (skin effect) localization mechanisms compete and coexist in complex networks. 

 By representing lattice sites as vertices and possible hopping paths as weighted edges, the graph formulation allows us to incorporate structural randomness and directional asymmetry into the model. In particular, the random graph approach offers a versatile way to study how stochastic connections influence the spectral and transport properties of the system.
In this study, we model the long-range connections using a directed Erdős–Rényi random graph \cite{frieze2015introduction}.

The primary motivation of this work is to investigate how localization phenomena evolve when a conventional one-dimensional lattice is extended into a random graph–like geometry. In such an extended configuration, the inclusion of random long-range connections transforms the simple 1D chain into a complex network.\\
To further enrich the dynamics, we introduce non-reciprocal (asymmetric) hopping, which inherently breaks Hermiticity and induces directed transport. This allows us to probe how localization is traditionally understood in the context of disorder. By systematically tuning the probability of long-range connections and the degree of non-reciprocity, we aim to uncover how directedness modifies the nature of localization, and the competing effects in them. We consider one tight-binding model for our study as it offers a wider representation of tight-binding models in condensed matter physics. 
Investigating these systems thus offers a unified platform to study non-Hermitian Anderson transitions, real–complex spectral phase boundaries, and the robustness of topological phenomena beyond translationally invariant lattices, bridging concepts from disordered quantum systems, random matrix theory, network science and other real world scenarios. 
\section*{Model Hamiltonian}
We consider a one-dimensional non-Hermitian SSH model consisting of $N$ sites, incorporating asymmetric hopping amplitudes, on-site disorder, and random directed long-range couplings. The inclusion of these directed long-range links effectively introduces the system with the characteristics of a random graph like lattice. The Hamiltonian is written as,
\begin{equation}
\begin{aligned}
H = \sum_{n=1}^{N/2} \Big[ \,
& t_{1,n}\, c_{2n-1}^\dagger c_{2n} 
+ t_{2}^{(+)}\, c_{2n}^\dagger c_{2n+1}  \\
& +\, t_{2}^{(-)}\, c_{2n+1}^\dagger c_{2n}
+ \text{H.c.} \,\Big]
+ H_{\text{LR}} .
\end{aligned}
\label{eq:H_main}
\end{equation}

where $c_i^\dagger$ ($c_i$) creates (annihilates) a particle at site $i$. The intra-cell hoppings are disordered as
\[
t_{1,n} = t_1^{(0)} + \delta t_{1,n}, \qquad \delta t_{1,n} \in [-W, W],
\]
and the inter-cell hoppings are non-reciprocal,
\begin{equation}
t_{2}^{(+)} = t_2 (1 + \Delta), 
\qquad 
t_{2}^{(-)} = t_2 (1 - \Delta),
\end{equation}
The asymmetry parameter $\Delta$ quantifies only the asymmetry in the system and it does not represent a superconducting gap, nor the maximum vertex degree as commonly denoted by $\Delta$ in graph theory. The system is Hermitian at $\Delta=0$ and becomes increasingly non-Hermitian as $\Delta$ grows.\\
The term $H_{\text{LR}}$ adds random long-range directed couplings between non-adjacent sites given as, 
\begin{equation}
H_{\text{LR}} = 
\sum_{i<j} \eta_{ij} \Big(
t_{ij}^{(+)} c_i^\dagger c_j
+ t_{ij}^{(-)} c_j^\dagger c_i
\Big).
\end{equation}

where $\eta_{ij}$ is a random variable equal to $1$ with probability $p$ and $0$ with probability $1-p$. This probability $p$ controls the density of long-range connections, transforming the 1D chain into a partially random network. For each active link, the hopping direction is chosen randomly, making the network directed,
\[
t_{ij}^{(+)} = t_2 (1 + \Delta), 
\qquad 
t_{ij}^{(-)} = t_2 (1 - \Delta).
\]

 In this model, each ordered pair of distinct sites $(i,j),$ $(i \neq j)$ is connected independently with probability $p$, allowing both directions $i \to j$ and $j \to i$ to exist simultaneously. For any such connection, the corresponding hopping amplitudes are $t_{ij}^{(+)} = t_2(1+\Delta)$ and $t_{ij}^{(-)} = t_2(1-\Delta)$, which serve as directed edge weights with a common base strength $t_2$ and a non-reciprocal asymmetry governed by $\Delta$. Hence, the parameter $p$ controls the connectivity of the random network, while $\Delta$ tunes its non-Hermiticity. 
\begin{figure}[H]
\centering
\includegraphics[width=0.8\columnwidth]{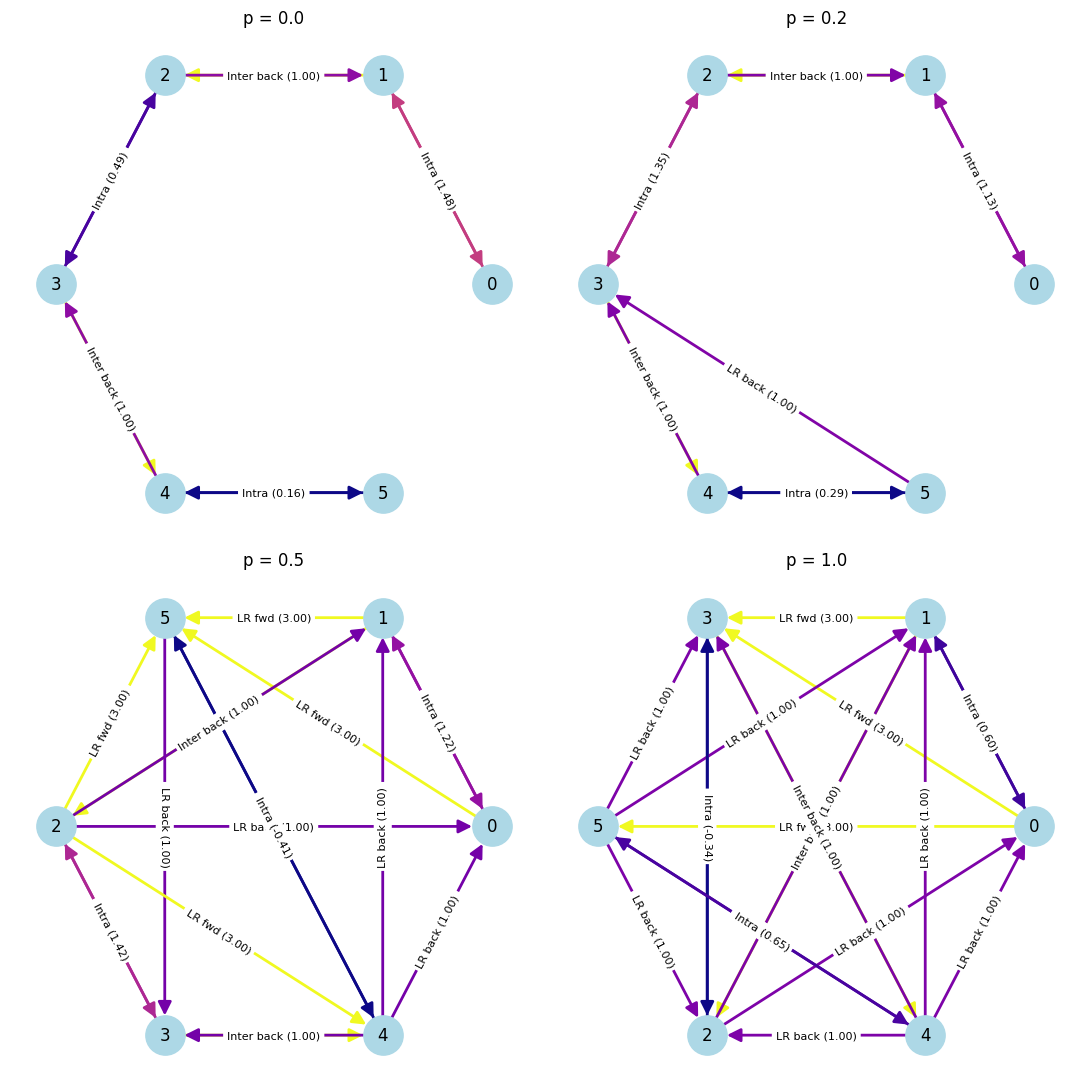}   
\caption{Schematic representation of random graph lattice for the range of values of directed long-range connections $p.$} 
\label{sch}
\end{figure}
\begin{figure}[H]
\centering
\includegraphics[width=1.0\columnwidth]{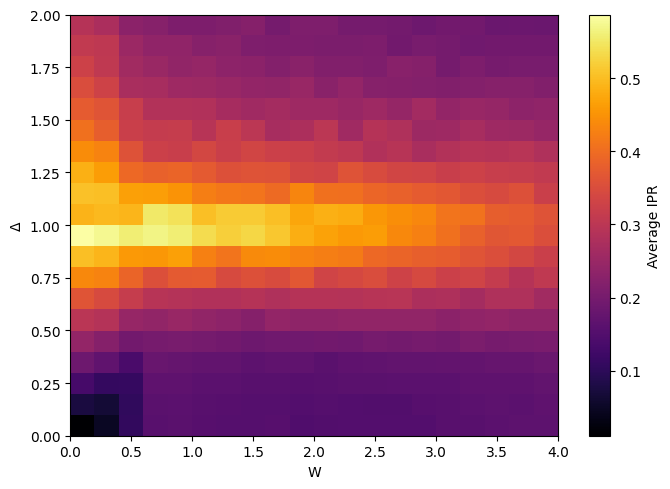}   
\caption{Phase diagram in the $(\Delta,W)$ plane represented in terms of the average IPR for $p=0$. The computation of average IPR is performed for the system size $N=300.$} 
\label{pd1}
\end{figure}
We begin our analysis by examining the phase diagram shown in Fig.~\ref{pd1} in $(\Delta, W)$ parameter space for $p=0$. We do this analysis using the measure of IPR. The IPR is a measure of spatial localization of eigenstates. For a normalized eigenstate $\psi_n$ defined over lattice sites L, the IPR is given by, 
\begin{equation}
\mathrm{IPR}_n = \sum_{j=1}^{L} \left| \psi_n(j) \right|^4 ,
\label{eq:ipr_def}
\end{equation}
where $\psi_n(j)$ denotes the amplitude of the $n$-th eigenstate at site $j$.
A high IPR is noticeable at $\Delta$=1 for a certain range of disorder values $W.$ 
\begin{figure}[H]
\centering
\includegraphics[width=1.0\columnwidth]{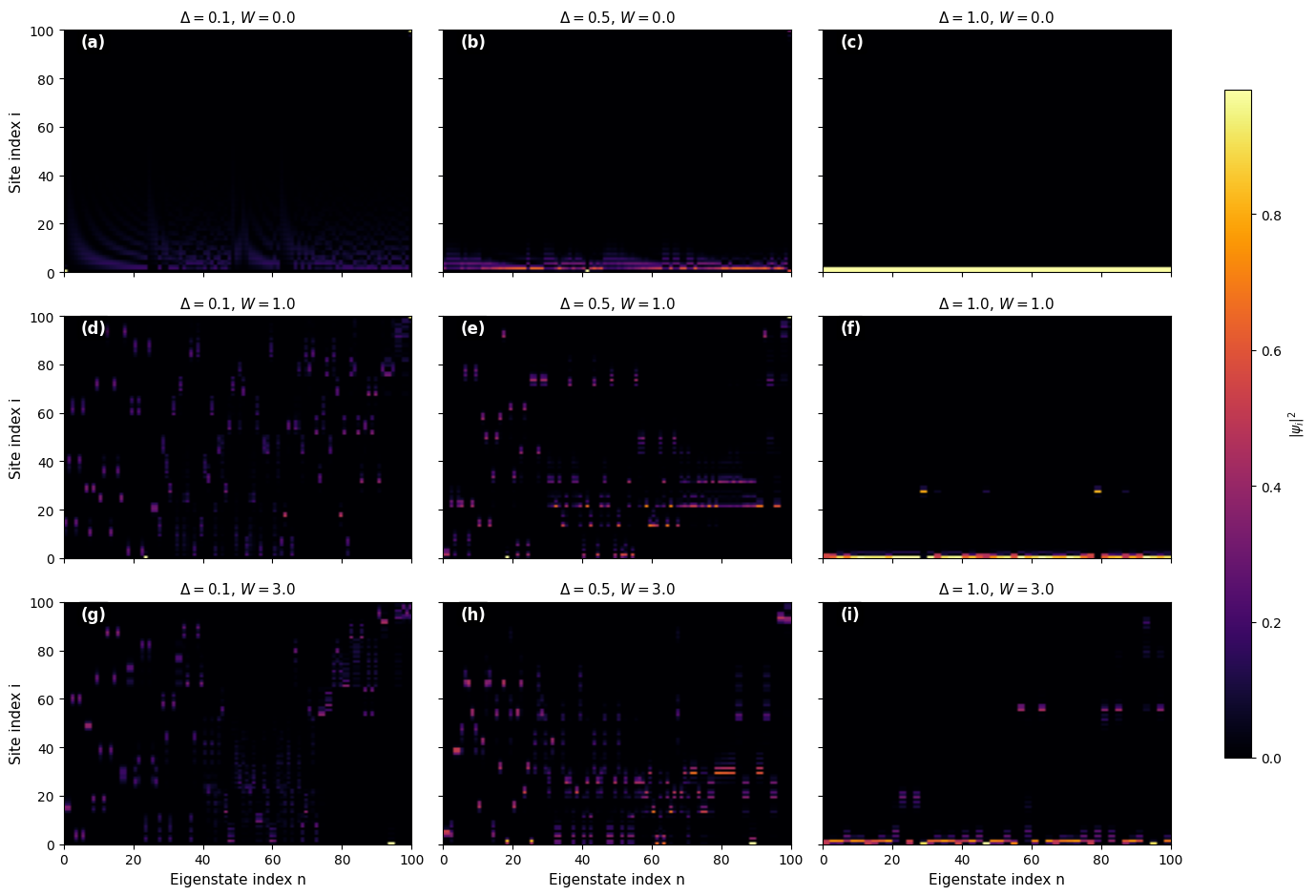}  
\caption{Spatial distribution of eigenstates with respect to site index in the absence of directed long-range connections $p$. Plots (a)-(c) are for clean limit $W=0$. (d)-(f) are plotted for $W=1.0$ and (g)-(i) are plotted for $W=3.0$ respectively.} 
\label{sch2}
\end{figure}
At $\Delta$=1 the backward hopping amplitude $t_{ij}^{(-)}$ vanishes, leaving only the forward hopping term $t_{ij}^{(+)}$. 
This fully unidirectional hopping configuration gives rise to the non-Hermitian skin effect, wherein the eigenstates become exponentially localized near one edge of the system. Consequently, the inverse participation ratio (IPR) exhibits a pronounced increase around $\Delta$=1, reflecting localization driven by skin effect. The phase diagram illustrated in the fig.\ref{pd1} is studied for higher values of $p$ which are presented in the appendix.\\ 
In connection to the phase diagram shown in the fig.\ref{pd1}, we study the probability distribution of eigenstates given by $\psi^2$ with respect to the asymmetric hopping $\Delta$ and on-site disorder $W$. Here the random directed long-range connections $p=0$, presented in the fig.\ref{sch2}.  
 Fig.\ref{sch2}(a)--(c) corresponds to the clean limit ($W=0$), where the eigenstates exhibit clear boundary accumulation with increasing $\Delta$, a hallmark of the non-Hermitian skin effect arising from asymmetric inter-cell couplings. As disorder is introduced, panels (d)--(i), the competition between asymmetric hopping and random on-site potential becomes evident.\\
Moderate disorder partially suppresses the skin effect by redistributing the eigenstates across the bulk, while stronger disorder, $W=3.0$ leads to conventional Anderson-type localization, characterized by randomly localized states throughout the lattice. Thus, the figure captures a smooth crossover from skin-dominated boundary localization in the clean asymmetric limit to disorder-induced bulk localization at higher $W$. Despite the strong on-site disorder, skin dominated boundary localization is still present as it can be seen from the panel (i).
\begin{figure}[H]
\centering
\includegraphics[width=1.0\columnwidth]{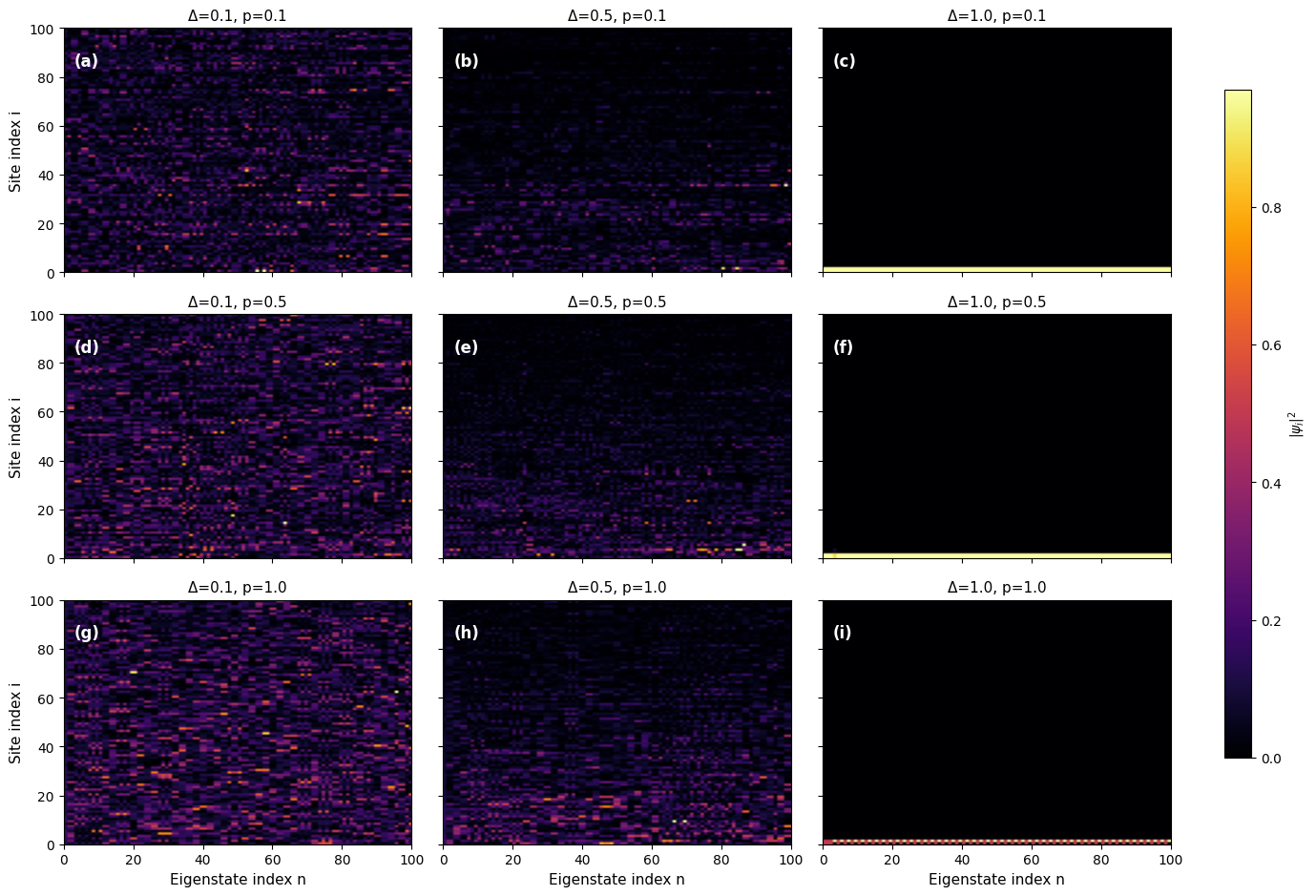}   
\caption{Spatial distribution of eigenstates with respect to site index in the absence of on-site disorder $W$. Plots (a)-(c) are for $p=0.1$. (d)-(f) are plotted for $p=0.5$ and (g)-(i) are plotted for $p=1.0$ respectively.} 
\label{dens1}
\end{figure}
In the clean limit $(W=0)$, the spatial distribution of eigenstates reflects the competition between asymmetric hopping and long-range connectivity. For small values of both $\Delta$ and $p$, the eigenstates remain relatively extended throughout the bulk, owing to the finite connectivity introduced by the random long-range couplings. As the asymmetric coupling $\Delta$ increases, the system exhibits a strong non-Hermitian skin effect, leading to a strong accumulation of eigenstates at one edge, as evident from Fig.~\ref{dens1}(a)–(c). With a gradual increase in the long-range coupling $p$, the eigenstates tend to spread more evenly across the bulk; however, the enhanced asymmetric hopping ultimately suppresses this delocalization as seen from the Fig.~\ref{dens1}(d)–(i). Although the skin effect–driven edge localization remains dominant at large $\Delta$, the increasing long-range coupling dilutes the strength of localization $\psi^2$, resulting in a partial redistribution of the eigenstates back into the bulk.

To gain deeper insight into this behavior, we further examine the variation of IPR as a function of the long-range coupling $p$ at a fixed value of $\Delta$. 

\begin{figure}[H]
\centering
\includegraphics[width=1.0\columnwidth]{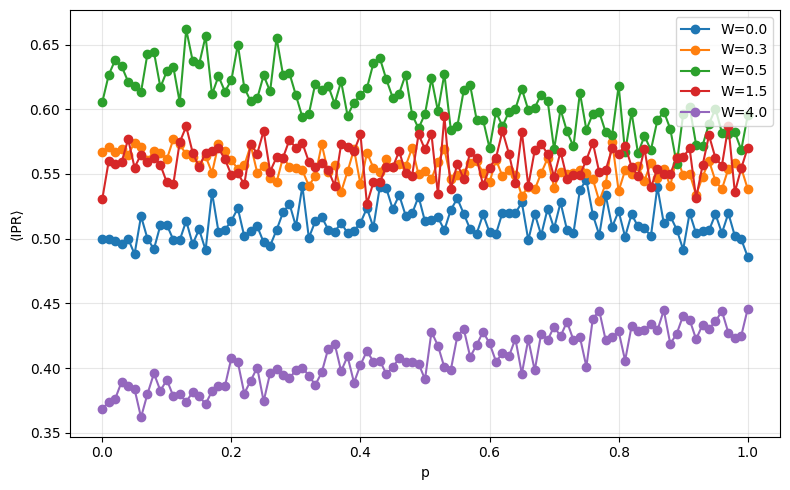}   
\caption{Average IPR is computed with respected to $p$ at $\Delta=1$ for different disorder strengths. The system size under consideration is $N=300.$ } 
\label{LSD1}
\end{figure}

In the Fig.\ref{LSD1}, we analyze the behavior of IPR with respect to $p$ for various disorder values at $\Delta=1$. For $W=0$, the IPR remains remains nearly constant around $0.5.$ This occurs because, in the clean limit at $\Delta=1$, the system already exhibits strong localization due to the non-Hermitian skin effect and this edge-localized modes are largely unaffected by the introduction of long-range directed couplings $p.$ However, interestingly for a low disorder case, the IPR shows a decreasing trend with increase in the values of $p.$ This occurs because the on-site disorder localizes states in the unit cell (Anderson localization) and the asymmetry in the coupling $\Delta$ pushes the states for edge localization driven by skin effect.
 In this competing setup, the introduction of $p$ opens up multiple directed pathways in the random lattice. These multiple pathways are utilized for localization driven by skin effect there by reducing the overall IPR scale. This trend is observed for several values of disorder $W=1.5,$ where the decreasing trend of the IPR transits to nearly constant. Further increasing the disorder, the IPR shows an increasing trend indicating an enhancement in localization. This enhancement in localization is driven by long-range random coupling as increase in $p$ creates multiple pathways competing with the stronger on-site disorder. Despite the strong on-site disorder, the availability of multiple channels increases the average IPR indicating the competing behavior of skin effect driven localization with Anderson-localization.

To distinguish if the localization is driven by Anderson type and skin effect, we make use of distribution of energy spectrum in both periodic boundary (PBC) and open boundary settings (OBC). The localization arising from skin effect would lose its localization property when the system is subjected PBC. In the Anderson type of localization, the localization remains broadened irrespective of PBC and OBC.   
\begin{figure}[H]
\centering
\includegraphics[width=0.9\columnwidth]{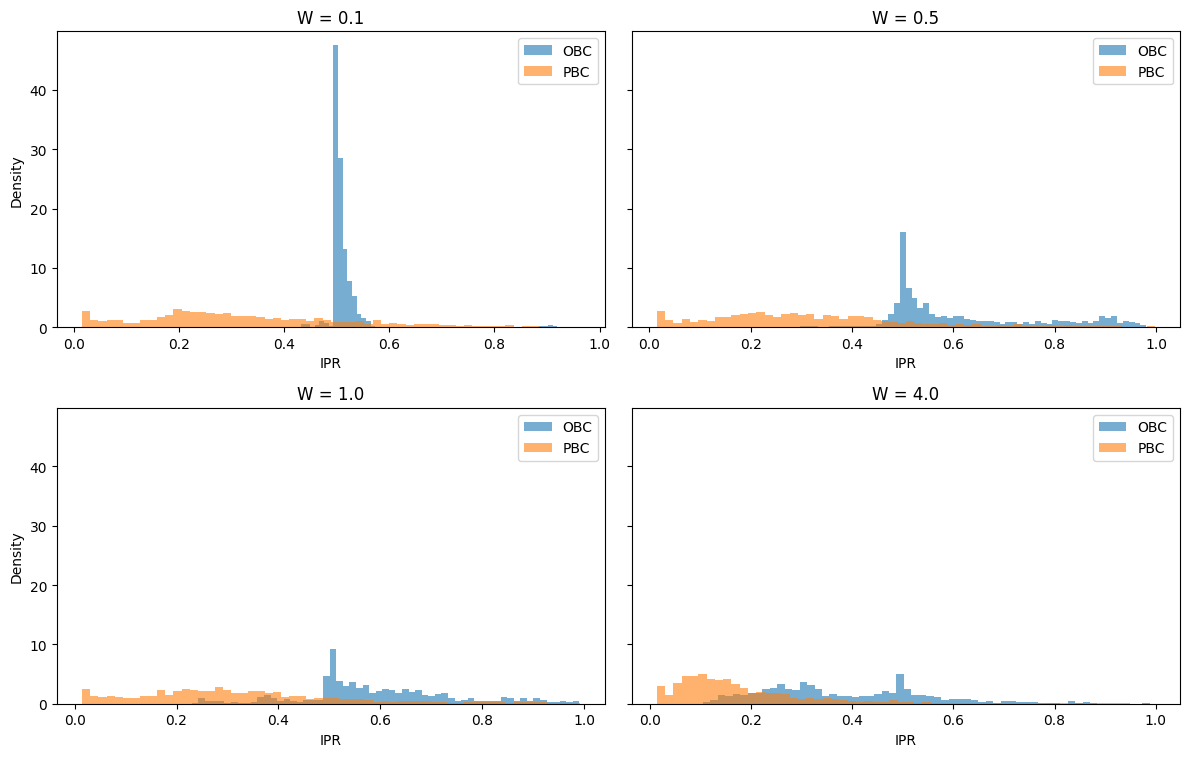}   
\caption{Density of eigenstates in both OBC and PBC are plotted with respect to the average IPR for $p=0.2.$ Different panels represent the distribution for different disorder strengths with the system size $N=300.$ } 
\label{Den1}
\end{figure}

Figure~\ref{Den1} compares the inverse participation ratio (IPR) distributions of eigenstates for OBC and PBC across increasing disorder strengths \( W \) in the non-Hermitian SSH-like model with asymmetric hopping (\( \Delta = 1.0 \)) and directed long-range connections (\( p = 0.2 \)). In the clean and weakly disordered regimes (\( W \lesssim 0.5 \)), the IPR under OBC exhibits a pronounced peak at high values, while the corresponding PBC states remain delocalized, indicating strong boundary-sensitive localization characteristic of the non-Hermitian skin effect. As the disorder strength increases (\( W \sim 1.0 \)), the distinction between OBC and PBC gradually diminishes, marking a crossover from skin effect driven localization to disorder-induced localization. At strong disorder (\( W = 4.0 \)), both boundary conditions yield similar, broadly distributed IPR spectra, consistent with Anderson driven localization that is insensitive to boundary conditions. This systematic evolution of the IPR distributions thus captures the transition from skin effect driven to disorder driven localization.
\begin{figure}[H]
\centering
\includegraphics[width=0.9\columnwidth]{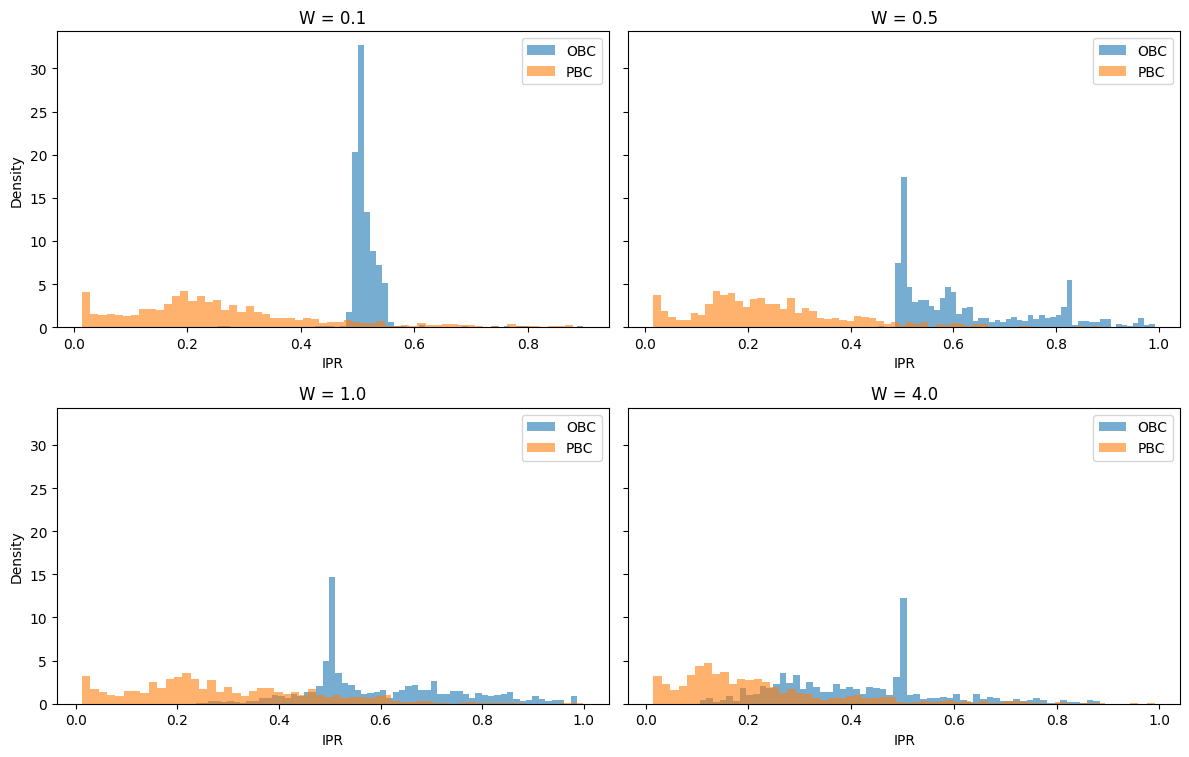}   
\caption{Density of eigenstates in both OBC and PBC are plotted with respect to the average IPR for $p=0.5.$ Different panels represent the distribution for different disorder strengths with the system size $N=300.$} 
\label{Den2}
\end{figure}

\begin{figure}[H]
\centering
\includegraphics[width=0.9\columnwidth]{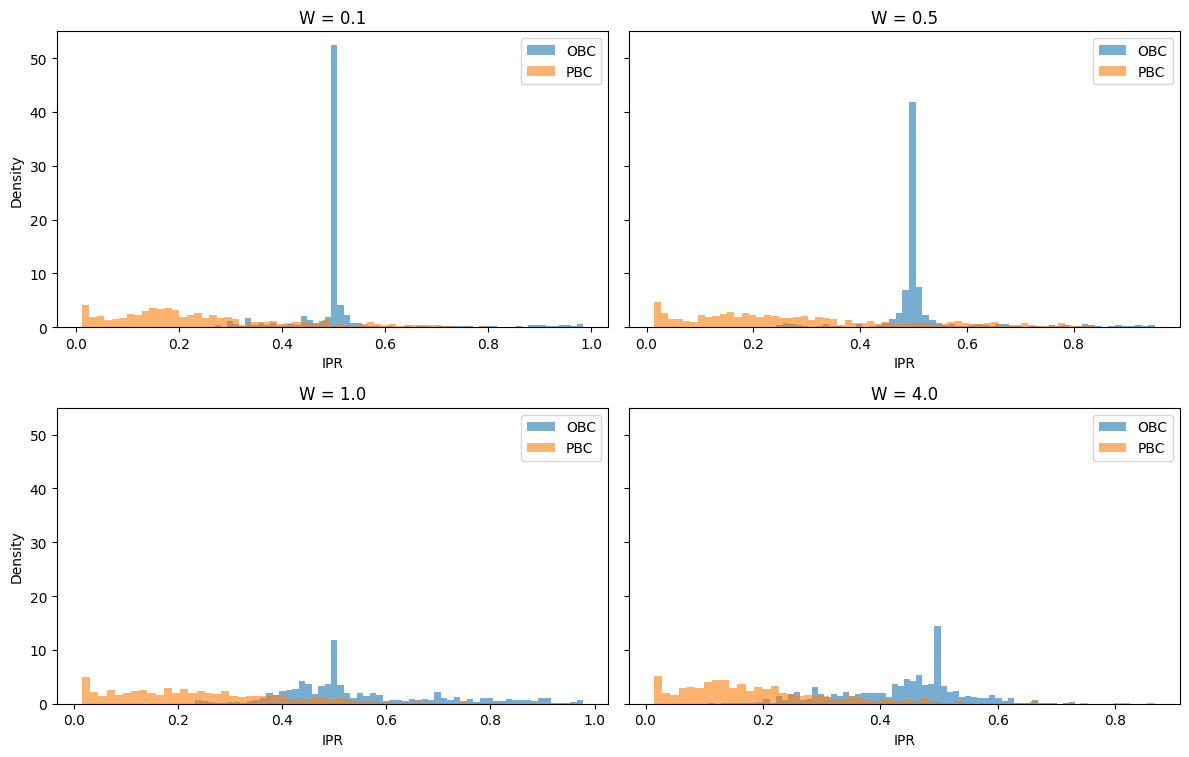}   
\caption{Density of eigenstates in both OBC and PBC are plotted with respect to the average IPR for $p=1.0.$ Different panels represent the distribution of the eigenstates for different disorder strengths with the system size $N=300.$} 
\label{Den3}
\end{figure}
Fig.\ref{Den2} and Fig.\ref{Den3} corresponds to IPR distributions for OBC and PBC for higher values of $p.$ At small $p,$ the system retains a predominantly one-dimensional character with sparse long-range links, leading to a pronounced asymmetry between OBC and PBC. Under OBC, the IPR distribution exhibits a sharp peak at large values, reflecting strong boundary localization associated with the non-Hermitian skin effect, whereas PBC yields a broader distribution centered at lower IPR, indicative of extended bulk states. Increasing $p$ to intermediate values $(p=0.5)$ introduces significant nonlocal connectivity that weakens the skin effect, producing flatter distributions and a reduced contrast between OBC and PBC. At full connectivity $(p=1.0)$, the two boundary conditions yield nearly identical IPR profiles, signaling that long-range hopping suppresses boundary accumulation and drives the system toward a bulk-dominated, random-network localization regime. Overall, the progressive similarity of OBC and PBC distributions with increasing $p$ highlights the crossover from boundary-induced to bulk-driven localization as the network becomes increasingly nonlocal.
\section*{Interplay of Skin Effect and Anderson Driven Localization Enhancement}
\begin{figure}[H]
\centering
\includegraphics[width=1.0\columnwidth]{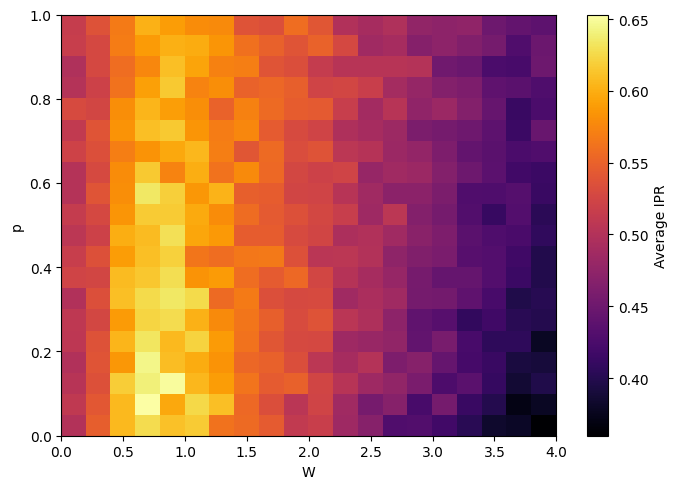}   
\caption{Phase diagram in the $(p,W)$ plane represented in terms of the average IPR at $\Delta=1$. The system size under consideration is $N=300.$} 
\label{gen}
\end{figure}

Fig.\ref{gen} illustrates the $(W, p)$ phase diagram at fixed $\Delta = 1$, thereby highlighting the role of directed long-range couplings. The high IPR ridge centered around moderate $W \approx 1$ corresponds to a regime of maximum localization, marking a crossover between skin effect to the left of the ridge and Anderson driven localization to the right. 
For the small disorder $W$ and a gradual increase in $p$, localization remains primarily driven by the skin effect, but becomes spatially redistributed due to multiple directed pathways introduced by random connections. At large $W$, disorder dominates and the localization becomes Anderson-like, resulting in a gradual reduction in the IPR. For a large disorder with increase in $p,$ the average IPR is increases because of the competitiveness of skin effect driven disorder.
At the ridge ranging from $W=0.5-1.5$, both skin effect driven and Anderson driven localization contributes to the enhancement of localization of states. In a large disorder limit, the states follow disorder driven localization but this does not rule out the fact that a small portion of skin effect driven localization is present as it can be seen from the Fig.\ref{Den1} - \ref{Den3}.
\section*{Relevance to the real world problems}
Although formulated in the physical context, the our model under consideration and its results are largely applicable in various fields including ML architecture, socio-economic status, migration.\\
Here we discuss few scenarios which are relevant for our model and results. 
The observed suppression of skin effect driven localization at low disorder with increasing non-local connectivity $p$ reflects how efficient the information can propagate, preventing bottlenecks and improving gradient flow across layers in ML architecture.\\
The competition between skin effect driven localization and Anderson driven localization provides a conceptual picture of how information flow can either concentrate within specific subnetworks or fragment due to excessive noise. In the ML context, skin effect driven localization is reminiscent of gradient accumulation in certain pathways, whereas Anderson localization corresponds to inactive or isolated subnetworks that fail to contribute to learning. Understanding this interplay offers a physics based perspective on issues such as vanishing gradients, or the emergence of inactive neurons in deep architectures.\\
The nodes of the network in our model can represent cities
or regions, while edges correspond to mobility pathways, trade links, or
social connections. Long-range couplings effectively model long-distance
migration channels or globalized economic interactions, whereas disorder acts as a proxy for local fluctuations in economic stability, employment
opportunities, or political conditions.
Asymmetric couplings introduce directional biases in the flow of people or
resources, analogous to migration driven by income gradients or the movement
of capital toward economically attractive hubs. Measures such as the IPR, used to quantify localization in physical systems, translate naturally to the concentration of population or economic activity in a few dominant regions. Depending on the interplay between long-range connectivity, disorder, and directional bias, the system may exhibit
localization, dispersal, or transition between these regimes, mirroring
realistic migration patterns, wealth concentration, and shock propagation in
economic networks.
\section*{Conclusion}
In summary, we have investigated the interplay of topology, disorder, and non-Hermiticity in a one-dimensional asymmetric lattice extended onto a generalized random graph. By incorporating asymmetric hopping, on-site disorder, and tunable long-range connections governed by a probability parameter $p,$ our study demonstrates how directed nonlocal coupling can modify the localization landscape of non-Hermitian systems.
 We show how the skin effect localization gets diffusive with introducing multiple random pathways. At $\Delta=1$, the coupling becomes unidirectional, and in the low-disorder regime, the IPR decreases with increasing $p,$ indicating a suppression of skin effect driven localization due to the introduction of multiple random pathways.
In contrast, in the high-disorder regime, we observe a crossover from skin effect dominated to Anderson driven localization, manifested as an increase in IPR with increasing $p,$ signifying the growing influence of disorder induced localization. 
 We show the competing interplay between skin effect driven and Anderson driven localizations which together shape the overall localization behavior of the system. This interplay and coexistence between skin effect driven and disorder induced localization result in enhancement of the overall localization of states in the system. 
  The random graph lattices themselves offer a powerful representation for reservoir computing and graph neural networks, where both randomness and spectral properties of the underlying connectivity strongly influence expressivity, memory capacity, and robustness. The tunable interplay of disorder, asymmetry, and nonlocality in our model therefore provides a versatile theoretical platform for designing and analyzing ML architectures with controlled information propagation, stability, and adaptability.
 The future scope of this study involves generalizing the results to other models and understanding the physical meaning of exceptional points in random graph lattices.  
 \onecolumngrid
\section*{Appendix}
\begin{figure}[H]
\begin{minipage}{0.30\linewidth}
\centering
\includegraphics[width=\linewidth]{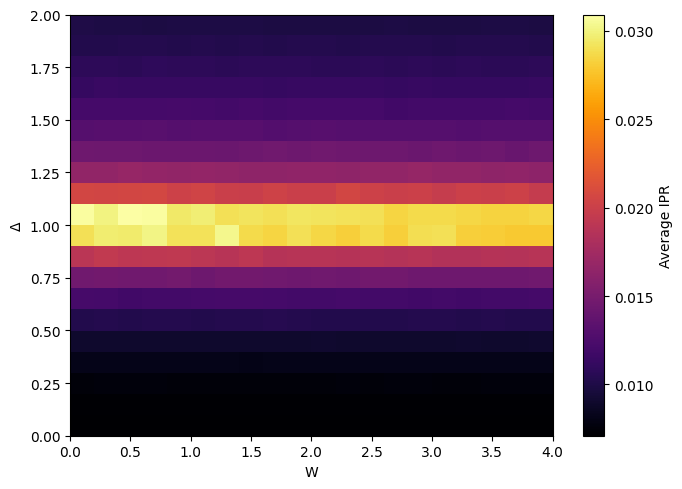}
\caption*{$p=0.1$}
\end{minipage}
\hfill
\begin{minipage}{0.30\linewidth}
\centering
\includegraphics[width=\linewidth]{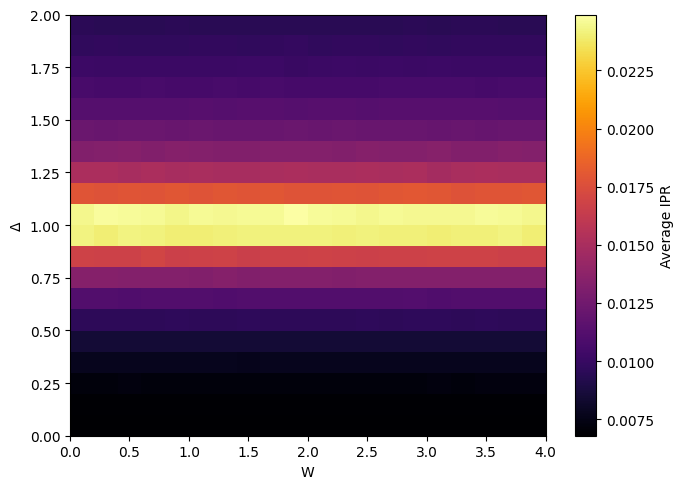}
\caption*{$p=0.5$}
\end{minipage}
\hfill
\begin{minipage}{0.30\linewidth}
\centering
\includegraphics[width=\linewidth]{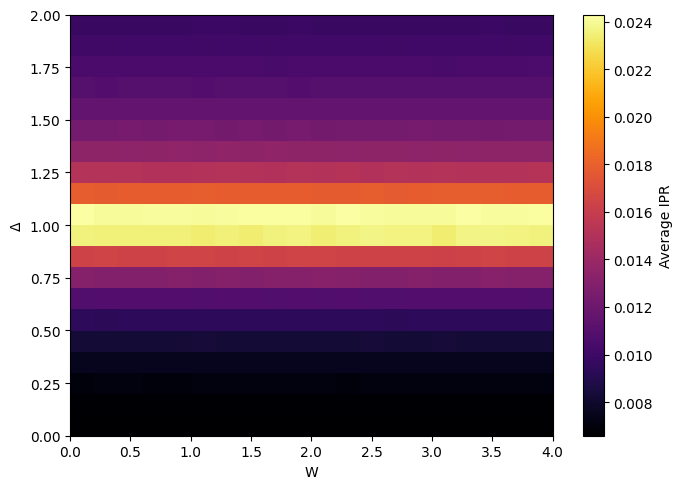}
\caption*{$p=1.0$}
\end{minipage}

 \caption{Phase diagram in the $(\Delta,W)$ plane presented in terms of the average IPR for different long-range connection strengths $p=0.1$, $0.5$, and $1.0$. The calculations are performed for a system size of $N=300$.}
    \label{ppd}
\end{figure}
\twocolumngrid

Fig.\ref{ppd} represents the phase diagram of asymmetric coupling $\Delta$ vs disorder $W$ for different values of $p$. The left, middle and right plots of Fig.\ref{ppd} corresponds to the values of $p,$ $0.1$, $0.5$ and $1.0$. As $p$ increases, the strength of the directed long-range couplings grows, extending the high IPR region around $\Delta=1$ for larger values of disorder $W.$ The persistence of this high IPR feature even at large disorder strengths indicates that localization is dominated by the highly connected random-graph structure, primarily driven by the non-Hermitian skin effect. With the introduction of finite $p$ the strong edge localization, characteristic of the skin effect remains, but becomes more diffusive leading to an overall reduction in the IPR scale. Physically, the presence of long-range directed links provides multiple alternative pathways for localization, causing the localized states to spread over a broader region instead of being confined to a single edge.
The high IPR region around $\Delta=1$ is particularly interesting in terms of its localization characteristics. 
 \bibliographystyle{unsrt}  
\bibliography{ref}

@article{PhysRevE.109.044315,
  title = {Localization transition in non-Hermitian systems depending on reciprocity and hopping asymmetry},
  author = {Kochergin, Daniil and Tiselko, Vasilii and Onuchin, Arsenii},
  journal = {Physical Review E},
  volume = {109},
  issue = {4},
  pages = {044315},
  numpages = {14},
  year = {2024},
  publisher = {American Physical Society},
  doi = {10.1103/PhysRevE.109.044315},
  url = {https://link.aps.org/doi/10.1103/PhysRevE.109.044315}
}

@article{yuce2016majorana,
  title={Majorana edge modes with gain and loss},
  author={Yuce, C},
  journal={Physical Review A},
  volume={93},
  number={6},
  pages={062130},
  year={2016},
  publisher={APS}
}

@article{wang2015spontaneous,
  title={Spontaneous PT-symmetry breaking in non-Hermitian Kitaev and extended Kitaev models},
  author={Wang, Xiaohui and Liu, Tingting and Xiong, Ye and Tong, Peiqing},
  journal={Physical Review A},
  volume={92},
  number={1},
  pages={012116},
  year={2015},
  publisher={APS}
}

@article{bergholtz2019exceptional,
  title={Exceptional Topology of Non-Hermitian Systems},
  author={Bergholtz, Emil J and Budich, Jan Carl and Kunst, Flore K},
  journal={arXiv preprint arXiv:1912.10048},
  year={2019}
}

@article{gong2018topological,
  title={Topological phases of non-Hermitian systems},
  author={Gong, Zongping and Ashida, Yuto and Kawabata, Kohei and Takasan, Kazuaki and Higashikawa, Sho and Ueda, Masahito},
  journal={Physical Review X},
  volume={8},
  number={3},
  pages={031079},
  year={2018},
  publisher={APS}
}

@article{lieu2018topological,
  title={Topological phases in the non-Hermitian Su-Schrieffer-Heeger model},
  author={Lieu, Simon},
  journal={Physical Review B},
  volume={97},
  number={4},
  pages={045106},
  year={2018},
  publisher={APS}
}

@article{klett2017relation,
  title={Relation between PT-symmetry breaking and topologically nontrivial phases in the Su-Schrieffer-Heeger and Kitaev models},
  author={Klett, Marcel and Cartarius, Holger and Dast, Dennis and Main, J{\"o}rg and Wunner, G{\"u}nter},
  journal={Physical Review A},
  volume={95},
  number={5},
  pages={053626},
  year={2017},
  publisher={APS}
}

@article{PhysRevB.31.2437,
  title = {Anderson localization in topologically disordered systems},
  author = {Logan, David E. and Wolynes, Peter G.},
  journal = {Phys. Rev. B},
  volume = {31},
  issue = {4},
  pages = {2437--2450},
  numpages = {0},
  year = {1985},
  publisher = {American Physical Society},
  doi = {10.1103/PhysRevB.31.2437},
  url = {https://link.aps.org/doi/10.1103/PhysRevB.31.2437}
}

@article{PhysRevLett.110.176403,
  title = {Topological Superconductor to Anderson Localization Transition in One-Dimensional Incommensurate Lattices},
  author = {Cai, Xiaoming and Lang, Li-Jun and Chen, Shu and Wang, Yupeng},
  journal = {Physical Review Letters},
  volume = {110},
  issue = {17},
  pages = {176403},
  numpages = {5},
  year = {2013},
  publisher = {American Physical Society},
  doi = {10.1103/PhysRevLett.110.176403},
  url = {https://link.aps.org/doi/10.1103/PhysRevLett.110.176403}
}

@article{Zeng_2022,
doi = {10.1088/1367-2630/ac61d0},
url = {https://doi.org/10.1088/1367-2630/ac61d0},
year = {2022},
publisher = {IOP Publishing},
volume = {24},
number = {4},
pages = {043023},
author = {Zeng, Qi-Bo and Lü, Rong},
title = {Real spectra, Anderson localization, and topological phases in one-dimensional quasireciprocal systems},
journal = {New Journal of Physics}
}

@article{PhysRevB.106.014207,
  title = {Interplay of disorder and point-gap topology: Chiral modes, localization, and non-Hermitian Anderson skin effect in one dimension},
  author = {Sarkar, Ronika and Hegde, Suraj S. and Narayan, Awadhesh},
  journal = {Physical Review B},
  volume = {106},
  issue = {1},
  pages = {014207},
  numpages = {12},
  year = {2022},
  publisher = {American Physical Society},
  doi = {10.1103/PhysRevB.106.014207},
  url = {https://link.aps.org/doi/10.1103/PhysRevB.106.014207}
}

@article{Wang_2021,
doi = {10.1088/2399-6528/ac261f},
url = {https://doi.org/10.1088/2399-6528/ac261f},
year = {2021},
publisher = {IOP Publishing},
volume = {5},
number = {9},
pages = {095011},
author = {Wang, R and Zhang, K L and Song, Z},
title = {Anderson localization induced by complex potential},
journal = {Journal of Physics Communications}
}

@article{estrada2013graph,
  title={Graph and network theory in physics},
  author={Estrada, Ernesto},
  journal={arXiv preprint arXiv:1302.4378},
  year={2013}
}

@article{PhysRevB.111.054116,
  title = {Graph theory based approach to identify phase transitions in condensed matter},
  author = {Wang, An and Sosso, Gabriele C.},
  journal = {Physical Review B},
  volume = {111},
  issue = {5},
  pages = {054116},
  numpages = {10},
  year = {2025},
  publisher = {American Physical Society},
  doi = {10.1103/PhysRevB.111.054116},
  url = {https://link.aps.org/doi/10.1103/PhysRevB.111.054116}
}

@article{PhysRevResearch.7.023132,
  title = {Graph-theoretical approach to the eigenvalue spectrum of perturbed higher-order exceptional points},
  author = {Grom, Daniel and Kullig, Julius and R\"ontgen, Malte and Wiersig, Jan},
  journal = {Physical Review Research},
  volume = {7},
  issue = {2},
  pages = {023132},
  numpages = {21},
  year = {2025},
  publisher = {American Physical Society},
  doi = {10.1103/PhysRevResearch.7.023132},
  url = {https://link.aps.org/doi/10.1103/PhysRevResearch.7.023132}
}

@article{kleftogiannis2024quantum,
  title={Quantum chaos, localization and phase transitions in random graphs},
  author={Kleftogiannis, Ioannis and Amanatidis, Ilias},
  journal={arXiv preprint arXiv:2412.14722},
  year={2024}
}

@book{frieze2015introduction,
  title={Introduction to random graphs},
  author={Frieze, Alan and Karo{\'n}ski, Micha{\l}},
  year={2015},
  publisher={Cambridge University Press}
}
 \end{document}